\documentclass[conference]{IEEEtran}

\usepackage{cite}
\usepackage{amsmath,amssymb,amsfonts}
\usepackage{algorithm, algorithmic}
\usepackage{graphicx}
\usepackage{textcomp}
\usepackage{xcolor}
\usepackage{acronym}
\usepackage[font=small]{caption}
\usepackage{subfig}
\usepackage{makecell}

\newacro{ai}[AI]{Artificial Intelligence}
\newacro{sota}[SOTA]{State-of-the-Art}
\newacro{snr}[SNR]{signal-to-noise ratio}
\newacro{psnr}[PSNR]{peak signal-to-noise ratio}
\newacro{ml}[ML]{machine learning}
\newacro{qos}[QoS]{Quality of Service}
\newacro{ae}[AE]{autoencoder}
\newacro{jscc}[JSCC]{joint source and channel coding}
\newacro{dnn}[DNN]{deep neural networks}
\newacro{csi}[CSI]{channel state information}
\newacro{cnn}[CNN]{convolutional neural network}
\newacro{rnn}[RNN]{recurrent neural network}
\newacro{scae}[SCAE]{simplicial convolutional autoencoder}
\newacro{csi}[CSI]{channel state information}

\def\BibTeX{{\rm B\kern-.05em{\sc i\kern-.025em b}\kern-.08em
    T\kern-.1667em\lower.7ex\hbox{E}\kern-.125emX}}
\begin{document}
\title{Joint Semantic-Native Communication and Inference via Minimal Simplicial Structures}

\author{
\IEEEauthorblockN{Qiyang Zhao\IEEEauthorrefmark{1}, Hang Zou\IEEEauthorrefmark{1}, Mehdi Bennis\IEEEauthorrefmark{2}, Mérouane Debbah\IEEEauthorrefmark{1}, Ebtesam Almazrouei\IEEEauthorrefmark{1}, Faouzi Bader\IEEEauthorrefmark{1}}
\IEEEauthorblockA{\IEEEauthorrefmark{1}Technology Innovation Institute, 9639 Masdar City, Abu Dhabi, United Arab Emirates}
\IEEEauthorblockA{\IEEEauthorrefmark{2}Centre for Wireless Communications, University of Oulu, Oulu 90014, Finland}
Email: \{qiyang.zhao, hang.zou\}@tii.ae, mehdi.bennis@oulu.fi, \\ \{merouane.debbah, ebtesam.almazrouei, carlos-faouzi.bader\}@tii.ae
}
\maketitle

\begin{abstract}
In this work, we study the problem of semantic communication and inference, in which a student agent (i.e. mobile device) queries a teacher agent (i.e. cloud sever) to generate higher-order data semantics living in a simplicial complex. Specifically, the teacher first maps its data into a k-order simplicial complex and learns its high-order correlations. For effective communication and inference, the teacher seeks minimally sufficient and invariant semantic structures prior to conveying information. These minimal simplicial structures are found via judiciously removing simplices selected by the Hodge Laplacians without compromising the inference query accuracy. Subsequently, the student locally runs its own set of queries based on a masked \ac{scae} leveraging both local and remote teacher's knowledge. Numerical results corroborate the effectiveness of the proposed approach in terms of improving inference query accuracy under different channel conditions and simplicial structures. Experiments on a coauthorship dataset show that removing simplices by ranking the Laplacian values yields a 85\% reduction in payload size without sacrificing accuracy. Joint semantic communication and inference by masked \ac{scae} improves query accuracy by 25\% compared to local student based query and 15\% compared to remote teacher based query. Finally, incorporating channel semantics is shown to effectively improve inference accuracy, notably at low \ac{snr} values. 
\end{abstract}

\vspace{1mm}
\begin{IEEEkeywords}
Semantic Communication, Semantic Inference, Simplicial Complex, Semantic Query
\end{IEEEkeywords}

\section{Introduction}
\label{sec:introduction}

Communication systems in the 6G era will ubiquitously connect intelligent agents, where the network natively supports communications between a plethora of \ac{ai} agents and models. Current \ac{sota} communication systems are based on Shannon's level A, which aims to accurately transfer and reconstruct information bits from a transmitter to receiver \cite{belfiore2022topos}. Under this paradigm, the network is oblivious to the information content being delivered and its effectiveness in solving tasks. Communicating large models or data as raw bits brings significant challenges to networks with limited capacity, energy, latency, etc. In contrast to this, transmitting semantic information enables higher communication efficiency without degrading system performance. Semantic information represents the underlying latent structure of information that is invariant to changes across data domains, distributions and context. Such structures should be minimal (in terms of size), yet efficient in performing targeted tasks. 

With the success of \ac{ml}, significant research works on semantic communications have emerged performing on extracting latent features from a given input, and communicating them to a receiver \cite{lu2022semanticsempowered}. For instance, the transformer architecture has shown big success in extracting semantic information from text messages, borrowing the bilingual evaluation understudy (BLEU) score as a semantic metric, compared to conventional source (Huffman) channel (Turbo) coding \cite{Xie2021}. In the domain of image transmission, \ac{cnn} has been applied incorporating channel noise into an autoencoder \cite{Bourtsoulatze}. 
Similarly, video transmission has been studied with contextual \ac{jscc} to optimize transmission rates \cite{Wang2023}. Accordingly, these works demonstrate an improved \ac{psnr} or structural similarity index (SSIM) of reconstructed images or videos at lower \ac{snr} or bandwidth. In the context of semantic channel coding, an adaptive universal transformer was proposed in \cite{Zhou2021}, by using \ac{csi} to adjust attention weights. While interesting, these works focus on learning latent representations directly from raw data, to compress data at the transmitter and reconstruct it at the receiver, without harnessing the structure of information. 

A different line of work casts the problem of semantic communication as a belief transport problem among teacher and student agents that reason over one another, sending only the minimum amount of semantic information \cite{Seo2023}. In \cite{Wang2021}, the authors model a knowledge graph of semantic symbols using attention based learning, to recover the transmitted text based on semantic similarity. Implicit semantics from graph representations have been studied in \cite{xiao2022reasoning}, using generative imitation based reasoning to interpret implicit relations between entities (symbols), offering reduced symbol error rates. A curriculum learning framework was developed in \cite{Farshbafan2023}, where a transmitter and receiver gradually identify the structure of the belief set as a description of observed events and take environment actions. Additionally, a neuro-symbolic \ac{ai} framework was studied in \cite{Thomas2022}, endowing nodes with reasoning-like capabilities.

A relevant scenario for studying semantic communication involves a student remotely learns a concept from a teacher via interaction. In this paper, instead of operating directly on raw data, we leverage semantic representations of data living on high dimensional topological spaces and make actionable decisions. We focus on how to effectively learn simplicial structures, with minimal semantic information communicated from the teacher to the student, to achieve specific goals. 

The main contribution of this paper are: 1) We model semantic information on high-order simplicial complexes, and minimize its structure using Hodge Laplacians across different dimensions; 2) We propose a masked \ac{scae} algorithm at the student agent to locally predict query data from its own semantic structure and the embedding received from the teacher; 3) We develop a joint semantic communication and inference scheme, leveraging both student and teacher agents' knowledge, to maximize communication efficiency and inference accuracy; 4) We incorporate channel semantics by training a \ac{scae} to improve system reliability at different channel conditions; and 5) We validate the proposed solutions on a coauthorship dataset by querying citations, showing how minimal semantic structures can be leveraged with improved query accuracy, notably at low \ac{snr} levels.

The rest of the paper is organized as follows. Section \ref{sec:system_model} introduces the system model of semantic communication on simplicial complexes. Section \ref{sec:resilient_communication} describes how to obtain minimal semantic structures and generating query data from both teacher and student's knowledge. Section \ref{sec:simulation} provides numerical results of the proposed approaches on a coauthorship dataset. Finally, the work is concluded with future work discussed in section \ref{sec:conclusion}. 
\begin{table}
    \vspace{2mm}
	\centering
	\caption{Notations}
	\label{tab:notations}
    \renewcommand{\arraystretch}{1.0}
    \resizebox{\linewidth}{!}{
	\begin{tabular}{|c|l|}
		\hline
		\textbf{Notation} & \textbf{Description} \\ \hline
        $S^K$ & $K$-order simplicial complex \\ \hline
        $\sigma^k_i $ & $i$-th $k$-order simplex, $k \in \left \{1,\dots,K \right \}$  \\ \hline
        $[L_k]_{i, j}$ & Hodge Laplacian at order $k$ between simplices $i$, $j$ \\ \hline
        $C^k$ & Set of $k$-cochains of simplicial complex $S$, $c \in C^k(S)$ \\ \hline
        $\pi$ & Linear coboundary operator \\ \hline
        $[B_k]_{i, j}$ & Coboundaries at order $k$ between simplices $i$, $j$ \\ \hline
        $\delta$ & Nonlinear activation function with bias $b$ \\ \hline
        $h$ & Hidden features of simplex cochains \\ \hline
        $\lambda_i^k$ & Eigenvalues of simplicial complex Laplacian matrix $L_k$ \\ \hline
        $f_e$, $f_g$ & Semantic encoder at transmitter and decoder at receiver \\ \hline
        $*$ & Simplicial convolution \\ \hline
        $D(\sigma,\mu)$ & number of edges between simplices $\sigma$ and $\mu$  \\ \hline
        $G(U, V, E)$ & \makecell[l]{Bipartite graph $G$ of independent sets $U$, $V$ and set of edges $E$} \\ \hline
        $\mathcal{H} (\sigma), \mathcal{T} (\sigma), \mathcal{G} (\sigma)$ & Known, received and generated set of simplices for query simplex $\sigma$ \\ \hline
	\end{tabular}
    }
    \vspace{-2mm}
\end{table}

\section{System Model and Scenario}
\label{sec:system_model}
The semantic communication problem under consideration is that of an interactive query in a teacher-student scenario, where a student leverages higher-order topological structures from its raw data and a few semantic data embedding from the remote teacher, to provide answers to a much larger set of queries. Specifically, the teacher uses a semantic encoder to extract minimal simplicial complexes. It describes the underlying structure of raw data, which is encoded as semantic embedding for transmission over a wireless channel. The student uses a semantic generator to predict query data from its local and remote teacher's semantic data embedding.

Concretely, we consider a query task of the coauthorship information \cite{Ebli2020}. The teacher learns semantic structures of the dataset on $k$-order simplicial complexes. Given a finite set of $K$ vertices $V = \left \{v_0,\dots,v_{K-1} \right \}, $ a $k$-simplex $\sigma^k$ is a topological object with $(k+1)$ vertices, containing $(k+1)$ faces of dimension $(k-1)$, denoted as $\{v_0,...,\hat{v}_i,...,v_k\}$. If a simplex $\sigma^k$ is a face of $\mu^{k+1}$, then $\mu^{k+1}$ is called a coface of $\sigma^k$. For a non-oriented simplex $\left \{v_0,\dots,v_{k} \right \}$, a possible oriented one is denoted as $\left [v_0,\dots,v_{K-1} \right ]$. 
A simplicial complex $S^K$ is thus a collection of simplices $\sigma^k$ with a dimension $k \in \left \{1, \dots, K \right \}$, that is closed under inclusion of all faces. Effectively, given our dataset we map a paper with $k$ authors into a $(k-1)$-simplex. With the paper's features (i.e. citations) projected on a simplicial complex, higher-order semantic information of the corpus can be extracted. A simplicial complex $S^K$ can be constructed from a bipartite graph $G = (U, V, E)$, where
two disjoint and independent sets of vertices $U$ and $V$ are connected through set of edges $E$. A $k$-simplex is any $k+1$ vertices in $V$ that has at least one common neighbour in $U$. A simplex $\sigma^{k} = \left \{ v_0, \dots, v_{k} \right \} \subseteq V$ can be defined as: 
\begin{equation}
	\label{eq:coboundary}
    \left\{\sigma^{k} \; | \; \exists u \in U \;\text{and}\; \exists \mu   \supseteq \sigma^{k}   \; \text{s.t.} \; (u,v_i) \in E, \forall v_i \in \mu \right\}
\end{equation}
where the weights on $U$ are assigned to the cochains of the projected simplices. We define $C^k(S)$ as the $k$-cochains of the simplicial complex $S$ (orders will be omitted if there is no confusion in the rests of paper) in an $\mathbb{R}$-vector space, which encodes raw data into a simplicial complex. The linear coboundary operator for a $k$-cochain $f$ can be defined as:
\begin{equation}
\label{eq:coboundary}
\begin{gathered}
\pi_k:C^k(S)\rightarrow C^{k+1}(S) \\
\pi_k(f)([v_0,...,v_{k+1}])=\sum_{i=0}^{k}(-1)^i f([v_0,...,\hat{v}_i,...,v_{k+1}])
\end{gathered}
\end{equation}

In the coauthorship use case, we refer to $U$ as papers and $V$ as authors. The citations of papers are associated with each vertex $u \in U$. The cochain function $c$ is the sum of citations of all the vertices $u$ (papers) that are connected to a set of vertices $\sigma^k$ (authors). The process of extracting the $(k-1)$-complex starts from the highest dimension by searching papers with $k$ authors, and applying the cochain function to the citations over connected vertices $u$. The process continues until the lowest dimension (classical graph). 

\begin{figure}[ht]
\centering
\includegraphics[width=1.0\columnwidth]{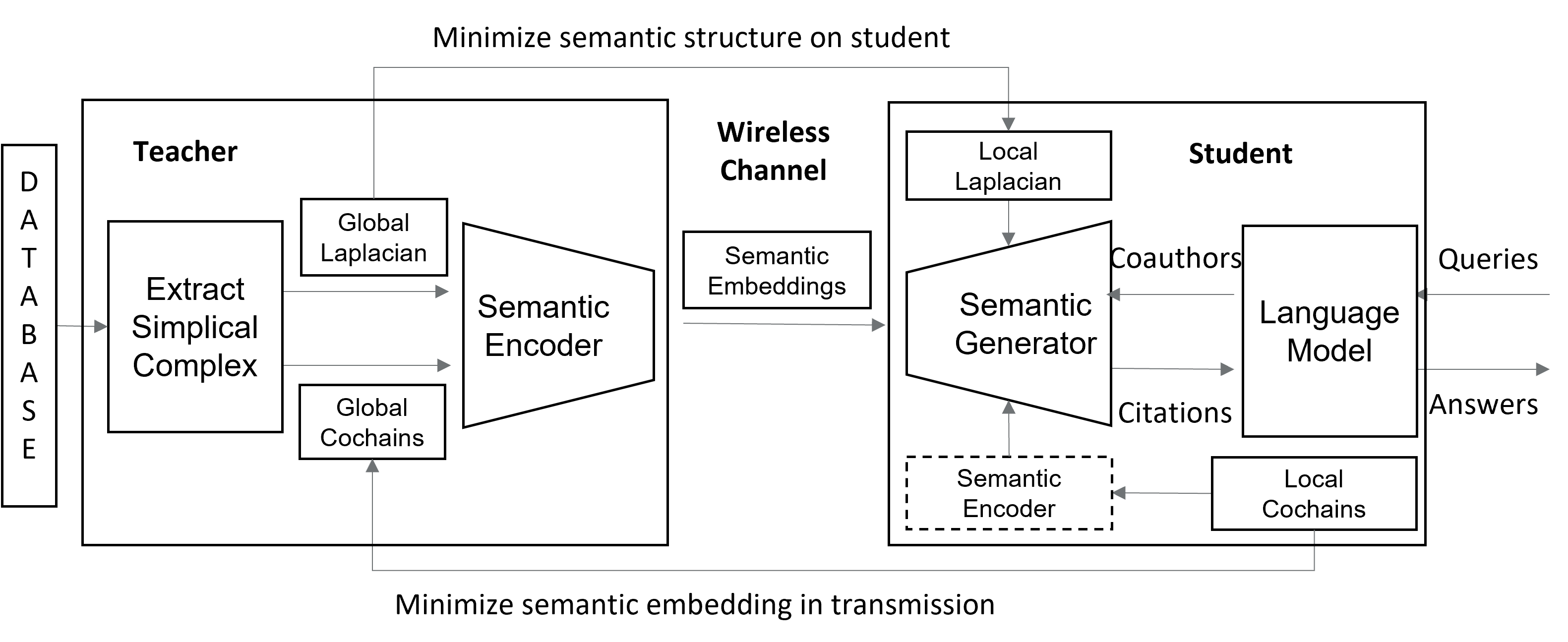}
\caption{System Model of Semantic Communication and Inference}\label{fig:paradigm} 
\vspace{-4mm}
\end{figure}

Query requests are generated at the student agent, asking about coauthor citations. The coauthorship database is stored at the teacher, where the simplicial complexes are extracted. A subset of the complex and cochain is sent to the student, to generate the query answer. 
We aimed to minimize the semantic structure at the student and semantic embedding at the teacher, and to maximize the accuracy of predicted citations. The system model is illustrated in Fig. \ref{fig:paradigm}. For general queries, a language model can be used to extract the simplices from queries, and produce answers of citations from cochains. 

\section{Semantic Communication and Inference on Minimal Simplicial Complexes}
\label{sec:resilient_communication}

\subsection{Minimal Semantic Structures on the Simplicial Complex}

The topological structure of a simplicial complex is determined by its set of vertices of the simplices. Following the bipartite graph relations, the cochains in a lower order simplex (e.g., $\sigma^1 = \left\{v_1, v_2 \right \}$) can be represented by  higher order simplices with a superset of its vertices (e.g., $\mu^2 = \left\{v_1, v_2, v_3 \right \}$). Moreover, two simplices with overlapping vertices (e.g., $\left\{v_1, v_2, v_3 \right \}$ and $\left\{v_0, v_2, v_3 \right \}$) have overlapping cochains. More generally, consider each simplex of dimension $k$ in coauthorship complex as a cochain query, the cochain function of the simplex $\sigma^k$ can be decomposed following the principle of inclusion-exclusion:  
\begin{equation}
\label{eq:cochain_simplex}
c(\sigma^k) = c_0 (\sigma^k) +  \sum_{k < \ell \leq K}{(-1)^{\ell - k + 1}} \sum_{\sigma^k 
 \subset\mu^\ell} c({\mu^\ell}),
\end{equation}
where $c_0 (\sigma^k)$ is independent citation of $\sigma^k$, i.e., citation of the paper authored uniquely by authors in $\sigma^k$.
Given a moment during the process of a sequence of queries, denote the set of all simplices to which $\sigma^k$ is subset with known cochains at student as $\mathcal{H} \left(\sigma^k \right)$, the iterative formulae to obtain the query can be written as: 
\begin{equation}
\label{eq:cochain_query}
\begin{gathered}
\begin{aligned}
c(\sigma^k) = & \underbrace{\sum_{k < l \leq K}{(-1)^{s - k + 1}} \sum_{ \mu^l \in \mathcal{H}\left(\sigma^k \right)}{c(\mu^l)}}_{\textrm{known}} \\ 
&  + \underbrace{ \sum_{k < l \leq K}{(-1)^{s - k + 1}} \sum_{ \mu^l \notin \mathcal{H}\left(\sigma^k \right)}{c(\mu^l)} + c_0 (\sigma^k)}_{\textrm{unknown}}
\end{aligned}
\end{gathered}
\end{equation}

In order to predict unknown information on simplices at different orders, we leverage the Hodge Laplacians to represent the simplicial complex structure. The $k$-th degree of the Hodge Laplacian $L_k$ contains two terms: a lower Laplacian $L_k^{\textrm{down}}$ and a upper Laplacian $L_k^{\textrm{up}}$, which encodes the lower and upper adjacency of $k$-order simplices, respectively. The higher order combinatorial Laplacian matrices are composed from lower and upper coboundary indices (denoting two simplices with the same orientation as $\sigma_i^{k - 1} \sim \sigma_j^{k}$, and with inclusivity property as $\sigma_i^{k - 1} \subset \sigma_j^{k}$, and vice versa). The $k$-Laplacian matrix can be obtained from incidence matrix $B$ as follows: 
\begin{equation}
    \label{eq:coboundaries}
    \begin{gathered}
    \begin{aligned}
    &L_k = B_k^TB_k + B_{k + 1}B_{k + 1}^T, k = 0,...,K-1
    \end{aligned}
    \end{gathered}
\end{equation}
\begin{equation}
    \label{eq:coboundaries}
    \begin{gathered}
    \begin{aligned}
    &[B_k]_{i, j} = \left\{
    \begin{aligned}
        0, \quad &\textrm{if} \;  \sigma_i^{k - 1} \not \subset \sigma_j^{k} \\
        1, \quad &\textrm{if} \; \sigma_i^{k - 1} \subset \sigma_j^{k} \; \textrm{and} \; \sigma_i^{k - 1}\sim \sigma_j^{k} \\
        -1, \quad &\textrm{if} \; \sigma_i^{k - 1} \subset \sigma_j^{k} \; \textrm{and} \; \sigma_i^{k - 1} \nsim \sigma_j^{k}
    \end{aligned}
    \right.
    \end{aligned}
    \end{gathered}
\end{equation}
The number of edges on the simplicial complex with non-zero Laplacians $L_k$ depends on the number of simplices pairs having a neighbour simplex at the $(k-1)$ order, and not belong to the same $(k+1)$ order simplex. In the bipartite graph, this means two simplices (coauthorhsips) have common vertices (authors), and not connecting to the same neighbour vertex $u$ (paper). This allows to predict or encode a cochain by using information passed from the adjacent ones, i.e., $	\left\{[L_k]_{i, j} \neq 0 \; | \; \sigma_i \cap \sigma_j  \neq \varnothing \right\}$. Simplicial convolution introduces local interactions between simplices via message passing. We invoke \cite{Ebli2020} and implement convolution filters restricted to low-degree polynomials in the frequency domain. We denote $h$ as the hidden feature of a cochain $c$, $W$ as trainable weights, $\psi$ as nonlinear activation function, and $\lambda_i^k$ as $i$-th eigenvalue of the Laplacian matrices $L_k$. The $N$th-degree simplicial convolutional layer can be defined as: 
\begin{equation}
	\label{eq:convolution_algebra}
	\begin{gathered}
	\psi \circ \left(h * \sum_{k = 0}^{N} W_i(\lambda_1^k, \lambda_2^k, ..., \lambda_{|\sigma^k|}^k) \right)
    \end{gathered}
\end{equation}

The filter is restricted to simplices that are within $N$ hops apart at the $k$-order. The complexity of inference is $O(|\sigma^k|)$, which scales with the number of $k$-order simplices. The complexity for training is fixed to $O(1)$. 

Specifically, for training the semantic encoder we apply a simplicial convolution to embed adjacent cochains into the semantic embedding, and in the semantic generator to predict missing cochains from adjacent known ones. The simplicial complex at higher order has a denser structure, because simplices have more overlapping vertices. In this case, a higher order Laplacian has larger impact on the simplical convolution, and accuracy of prediction. 

Our goal is to maximize the query accuracy under minimal simplicial structure represented by the Laplacian matrix. In a simplified single layer convolution, the problem can be written as the minimizing the non-zero elements of each $k$-order Laplacian via a binary matrix $b_k$. We assume that the weights $W$ are trained under the full structure to approximate the ground-truth cochains $c$ (under minimum error $\epsilon$), and the structure minimization problem can be written as: 
\begin{equation}
	\label{eq:maxmin}
    \begin{aligned}
    \min_{b_k} \; & \sum_{i = 0}^{|\sigma^k|} \sum_{j = 0}^{|\sigma^k|}{[b_k]_{i, j}\left|[L_k]_{i, j}\right|} \\
    \textrm{s.t.} \; &\left\| \psi \circ \left(\sum_{i = 0}^{|\sigma^k|}{W_i b_k L_k^i c}\right) - c \right\| < \epsilon \\
    & [b_k]_{i, j} \in \{0, 1\},\; \forall i,j
    \end{aligned}
\end{equation}
where $L_k^i$ denotes the $i$-th power of Laplacian matrix $L_k$.
According to (\ref{eq:convolution_algebra}), an edge with higher Laplacian value $[L_k]_{i, j}$ has a larger impact on the output of the simplicial convolution, thus decreasing the prediction accuracy. Similarly, a simplex $\sigma^k_i$ with higher degree value $[L_k]_{i, i}$ is affected more by message passing from others during convolution. We thus propose to minimize the simplicial structure by reducing the edges and simplices according to the Laplacian threshold $l$: 
\begin{itemize}
    \item \textit{Edge Minimization}: edges connecting simplices from different dimensions are ranked by the normalized Laplacian values $L_{i, j}$, and lower ranked edges are removed:
    \begin{equation}
    	\label{eq:maxmin}
        \left\{ [b_k]_{i, j} \leftarrow 0 | [L_k]_{i, j} \leq l, i \neq j \right\}
    \end{equation}
    \item \textit{Simplex Minimization}: simplices $\sigma^k$ from dimension $k$ are ranked by their degree values $L_{\sigma^k_i, \sigma^k_i}$. The lower ranked simplices and connected edges are removed.
    \begin{equation}
    	\label{eq:maxmin}
        \left\{ [b_k]_{i, j}, [b_k]_{j, i} \leftarrow 0 | [L_k]_{\sigma^k_i, \sigma^k_i} \leq l, j = 0\dots|\sigma^k| \right\}
    \end{equation}    
\end{itemize}

We will numerically evaluate the impact of different levels of $l$ on the structure's size and prediction accuracy.

\subsection{Masked Simplicial Convolutional Autoencoder}

Simplicial convolution can be used to impute the missing cochain data. The framework in \cite{Ebli2020} randomly masks the simplices with mean value of the cochains in the same dimension, and trains a convolutional model to minimize the $\mathcal{L}_1$-norm over known cochains. In our earlier work \cite{Zhao2022}, we applied this method to recover distorted data. In this work, we aim to further leverage the simplical complex structure to predict and generate data. Masked language models were used in pre-training BERT, by predicting the randomly masked tokens in a sentence \cite{devlin2019bert}. Meanwhile, masked image modeling is used on \ac{cnn} for BERT-sytle pre-training \cite{tian2023designing}. Both approaches shows effective data abstraction and generation from context. Inspired by these approaches, we develop a masked training and recursive prediction method on the simplicial complexes.

The \ac{scae} model contains a semantic encoder $f_e$ which encodes cochains into a latent embedding $z$, and a semantic generator $f_g$ which generates unknown cochains. Multiple layers of simplicial convolution are used on $f_e$ and $f_g$.
\begin{equation}
	\label{eq:maxmin}
    \begin{aligned}
    &z = f_e(c | L, W_e) \\
    &\hat{c} = f_g(z | L, W_g)
    \end{aligned}
\end{equation}

We start with the complete simplicial complex $S$ extracted from the training dataset. First, a set of large number $T$ of small simplical complexes $\mathcal{S}_{\text{train}} = \left\{{S^{(1)},\dots,S^{(T)}} \right\}$ at different orders are downsampled, where $\mathcal{S}_{\text{train}} \subset S$ and $T \ll |S|$. In the training phase, we randomly mask $p\%$ simplices with random cochain values. The masked simplices are within $N$ hops from known simplices. This is to guarantee message passing when using convolution. We train the model with an objective function to minimize the predicted masked cochains to their ground-truth citations. Denoting $ \mathcal{T} = \left\{\sigma_1, \dots,\sigma_T  \right\}$ as the set of masked simplices with $\sigma_i \in S^{(i)}$, $D$ as number of edges between simplices and $c_{\max}$ the maximum of coauthorship citation, the training procedure is formulated as: 
\begin{equation}
\label{eq:mask_scae}
\begin{aligned}
\min_{W_e, W_g} \; &   \sum_ {i=1} ^{T} | f_g(f_e(c(\sigma_i)) - \hat{c}(\sigma_i)  | \\
\textrm{s.t.} \; & c(\sigma_i) \leftarrow  U(1, c_{\max}),\; \forall i \\
& D(\sigma_i, S^{(i)}\backslash \sigma_i) \leq N,\; \forall i     
\end{aligned}
\end{equation}
where $U\left(a,b \right)$ means a random integer between $a$ and $b$. Different from the cochain imputation work \cite{Ebli2020}, our objective is to minimize the loss of the masked cochains instead of the known cochains.

\begin{figure}[ht]
\vspace{-3mm}
\centering
\includegraphics[width=0.9\columnwidth]{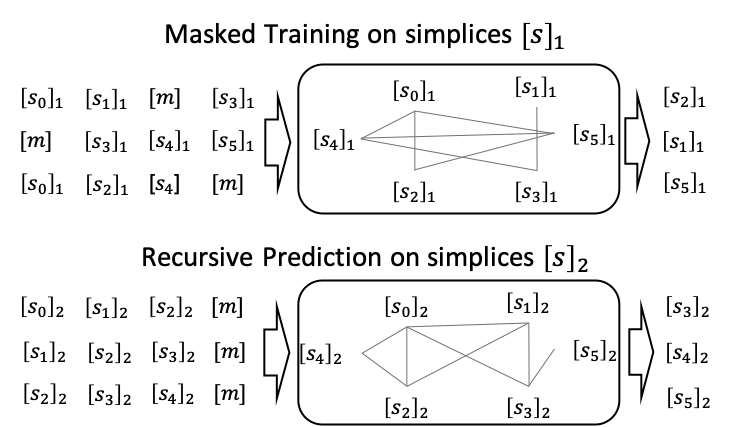}
\caption{Example of masked training and recursive prediction}\label{fig:model} 
\end{figure}

We leverage a recursive prediction to generate the missing citations. The missing simplices in $\mathcal{M}$ are ranked according to their highest Laplacian w.r.t. known simplices. It then selects maximum $p\%$ simplices to predict their cochains from the trained \ac{scae}. In the next iteration, the predicted simplices are added to the known simplices, to predict the next $p\%$ simplices based on the same selection criteria. The process continues until all queried citations are predicted. An example of the masked training and recursive prediction is given in Fig. \ref{fig:model}.
\vspace{-3mm}

\subsection{Joint Semantic Communication and Inference}

The masked \ac{scae} provides a method to predict cochains on the simplicial complex. To best leverage the knowledge of the student and teacher agent for query, we develop a joint semantic communication and inference scheme (Algorithm \ref{alg:comms_infer}). 

When the student receives a query request for coauthorship $\sigma_q$, it first encodes the known simplices $\mathcal{H}(\sigma_q)$ that are supersets of $\sigma_q$ with a semantic encoder $f_e(\mathcal{H} (\sigma_q))$. It then ranks the elements of $\mathcal{H}(\sigma_q)$ according to their highest Laplacians to $\sigma_q$, and downselects a set of simplices $\mathcal{G}(\sigma_q) \subset \mathcal{H}(\sigma_q)  $ with $|\mathcal{G}(\sigma_q)| = p|\mathcal{H}(\sigma_q)|$. In turn, the teacher encodes the remaining simplices $\mathcal{T}(\sigma_q) = \mathcal{H}(\sigma_q) \backslash \mathcal{G}(\sigma_q) $ and transmits the embedding $f_e(\mathcal{T}(\sigma_q)) $ to the student over wireless channel. The student then uses the semantic generator with the embeddings of known and received simplices to predict $\hat{c}(\sigma_q) = f_g(f_e(\mathcal{H}(\sigma_q) , \mathcal{T}(\sigma_q) ))$. Before ending the query, we update $\mathcal{H}$ based on $\mathcal{T}(\sigma_q)$ for every simplex. 
\begin{algorithm}
\caption{Joint Semantic Communication and Inference}
\label{alg:comms_infer}
\begin{algorithmic}[1]
\STATE \textbf{When} received query to coauthorship data $\sigma_q$, \textbf{do}:
\STATE Student encodes known simplices $f_e(\mathcal{H}(\sigma_q))$;
\STATE Student ranks the known simplices $\mathcal{H}(\sigma_q)$ w.r.t. $\sigma_q$ according to their highest Laplacians;
\STATE Student downselects simplices $\mathcal{G}(\sigma_q) \subset \mathcal{H}(\sigma_q)$ according to step (3) with $\mathcal{G} (\sigma_q)= p|\mathcal{H} (\sigma_q)|$;
\STATE Student requests teacher to transmit the cochains of simplices in $\mathcal{T}(\sigma_q) = \mathcal{H}(\sigma_q) \backslash \mathcal{G}(\sigma_q)$ if $\mathcal{T}(\sigma_q) \neq \varnothing$;
\STATE Teacher encodes $f_e(\mathcal{T}(\sigma_q))$ and send to student;
\STATE Student predicts $\hat{c}(\sigma_q)$ from $f_g(f_e(\mathcal{H}(\sigma_q), \mathcal{T}(\sigma_q)))$;
\STATE Update $\mathcal{H}(\sigma)$ for all simplex $\sigma \in S$;
\end{algorithmic} 
\end{algorithm}

In order to improve the performance in dynamic wireless channels, we incorporate channel information into the embedded cochains, and train a semantic generator to reduce the distortions. This enables the model to perform robustly under different channel conditions, without increasing the size of the structure at the student or sending more semantic embeddings.

\section{Simulation and Performance Evaluation}
\label{sec:simulation}
To corroborate the effectiveness of proposed approaches, we use a dataset from the Semantic Scholar Open Research Corpus \cite{Ammar2018}, which contains over 39 million research papers in computer science, with attributes such as author list and citations. To construct a co-authorship complex, we perform a random walk for $80$ papers with citations between 1 and 10, from a randomly sampled starting paper. A set of $k$-simplicial complexes is created by joining $k$-cochains sharing authors. An additive white Gaussian noise (AWGN) channel is used.

\begin{figure}[h]
\vspace{-2mm}
\centering
\includegraphics[width=0.9\columnwidth]{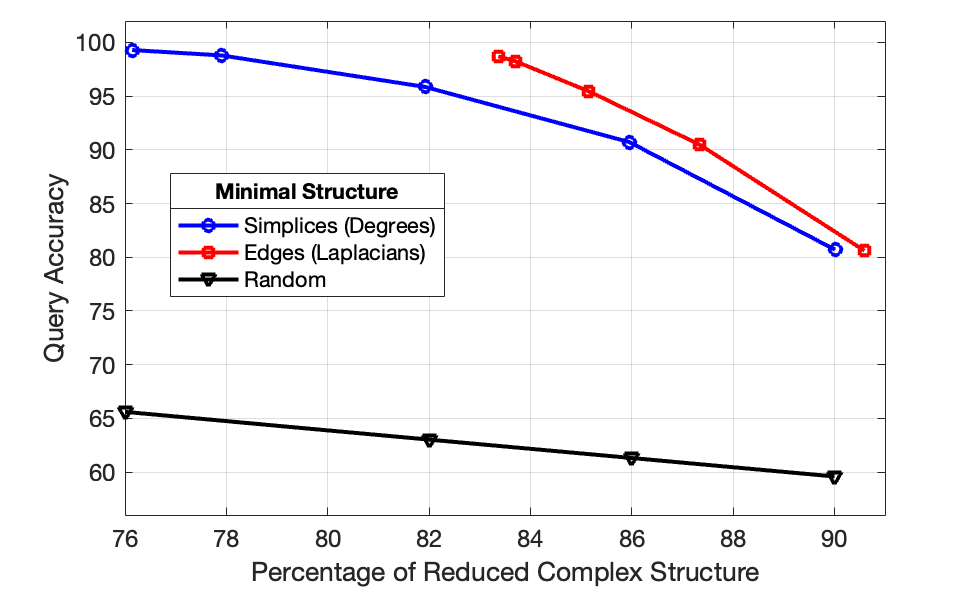}
\caption{Query accuracy under different percentages of reduced simplicial complexes}\label{fig:minimize_simplex}
\vspace{-2mm}
\end{figure}

We first evaluate the performance when minimizing the simplicial complex structure, using Laplacian or degree values to reduce edges or simplices. Fig. \ref{fig:minimize_simplex} demonstrates the query accuracy after applying the structure minimization schemes. First, we can see that reducing simplices by ranking with degrees significantly reduces the size of simplicial complexes by 76\%, without reducing query accuracy. Meanwhile, the accuracy starts dropping only after reducing 86\% of the complex size. Compared to a baseline scheme that randomly removes edges, our scheme improves accuracy by up to 35\%. Furthermore, when reducing edges using the Laplacian, the system performs slightly better and the accuracy starts dropping after only 83\% of the complex is reduced. This is because simplices with sparse edges can still be predicted from an edge with a high Laplacian value. In summary, a small percentage of simplices and edges with higher Laplacians have major impact on the query accuracy.

\begin{figure}[!h]
\vspace{-4mm}
\centering
\includegraphics[width=0.9\columnwidth]{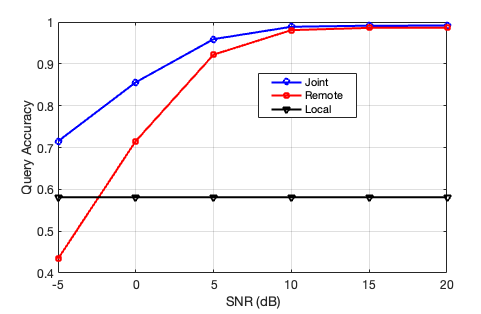}
\caption{Query accuracy of semantic communication and inference at different \ac{snr} levels}\label{fig:comms_infer}
\vspace{-8mm}
\end{figure}

\begin{figure}[!h]
\centering
\includegraphics[width=0.9\columnwidth]{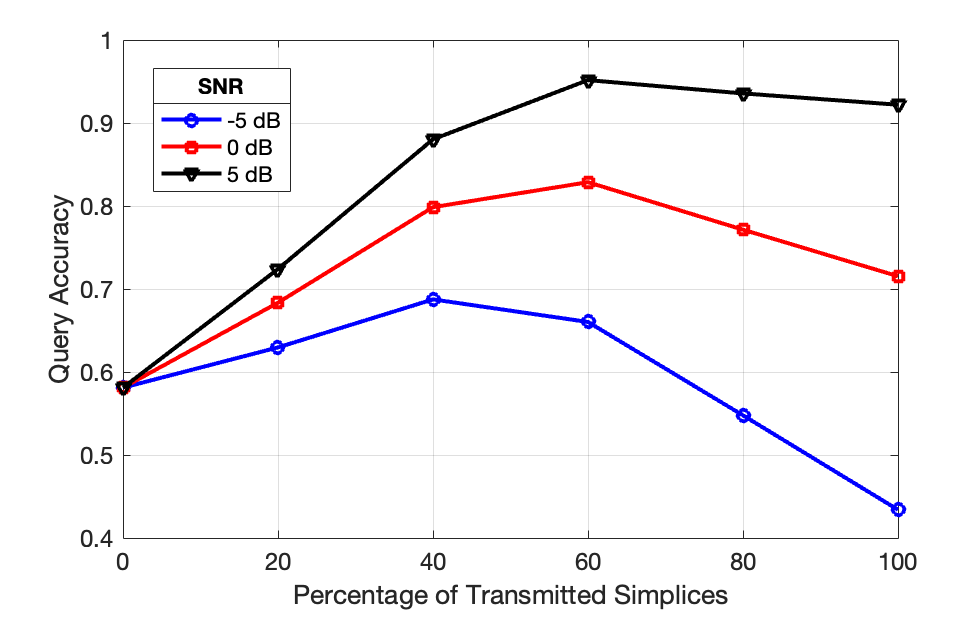}
\caption{Query accuracy under different percentages of transmitted simplices}\label{fig:comms_infer_percent} 
\end{figure}

In Fig. \ref{fig:comms_infer}, we evaluate the the joint semantic communication and inference based on a masked \ac{scae} with impact of \ac{snr}. The local query achieves 60\% accuracy when the student has 50\% of the known cochains. When fully relying on a query from the teacher, the accuracy is lower at $\ac{snr}=-5$ dB due to channel noise, and increases gradually as the channel improves reaching 100\% at $\ac{snr}=10$ dB. With joint semantic communication and inference, the accuracy is much higher at $\ac{snr}<5$ dB than both remote and local query. Furthermore, the query accuracy versus the percentage of semantic embedding transmitted from the teacher is shown in Fig. \ref{fig:comms_infer_percent}. It can be seen that at high $\ac{snr}>5$ dB, transmitting more information (60\%) achieves the best performance, whereas at low $\ac{snr}<-5$ dB, the best percentage of transmission decreases down to 40\%. These results indicate that jointly utilizing the student and teacher's knowledge can largely improve the query performance at poor channel conditions while reducing transmitted information.

\begin{figure}[!h]
\centering
\includegraphics[width=0.9\columnwidth]{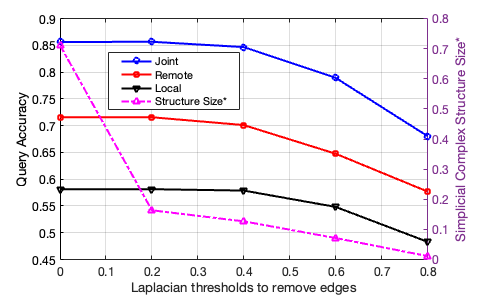}
\caption{Performance of semantic communication and inference under different percentages of minimal structure}\label{fig:masked_laplacian_accuracy} 
\end{figure}

\begin{figure}[!h]
\centering
\includegraphics[width=0.9\columnwidth]{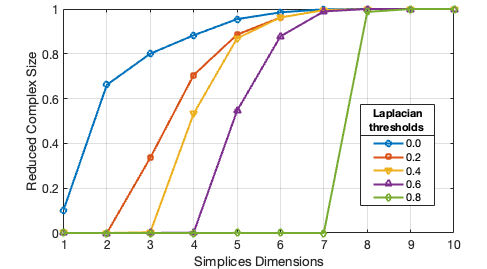}
\caption{Reduced simplicial complex size at each dimension}\label{fig:edge_laplacians}
\vspace{-3mm}
\end{figure}

The performance of combining minimal semantic structures with joint semantic communication and inference is shown in Fig. \ref{fig:masked_laplacian_accuracy}. First we can see that by reducing the simplicial complex size to 15\%, the accuracy of all the transmission schemes are not affected. Joint communication and inference has higher accuracy of 15\% higher than remote query and 25\% higher than local prediction. The performance drops by 30\% when the structure size is further reduced. We can see that with the edges removed by Laplacian thresholds above 0.4, the generation accuracy drops significantly. Fig. \ref{fig:edge_laplacians} shows the simplical complex size in different dimensions. We can see that with Laplacian threshold 0.4 all the edges on simplices below order 2 are removed, while those above order 7 are retained. This further demonstrates that a high order structure contains more semantic information, which is essential in achieving the targeted goal. This suggests the semantic structure can be minimized by removing most of the low order simplices.

\begin{figure}[h]
\vspace{-2mm}
\centering
\includegraphics[width=0.9\columnwidth]{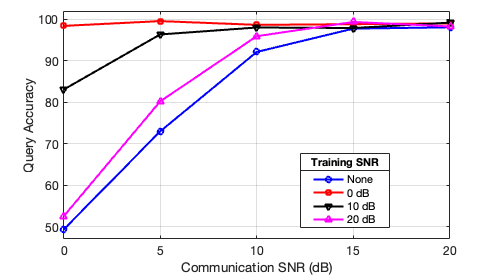}
\caption{Performance of training \ac{scae} with channel state information}\label{fig:channel_semantics} 
\vspace{-2mm}
\end{figure}

The performance of taking into account \ac{csi} is demonstrated in Fig. \ref{fig:channel_semantics} under the impact of \ac{snr}. It can be observed that training the \ac{scae} with low \ac{snr}, the system achieves significantly higher accuracy in communication at low \ac{snr}. The performance drops when the \ac{snr} is lower than that used in \ac{csi} for training. With training on \ac{snr} above 20 dB the accuracy is close to the baseline without \ac{csi}. This emphasizes that joint semantic channel coding significantly improves system reliability in poor channel conditions. 

\section{Conclusion}
\label{sec:conclusion}
In this paper, we proposed a joint semantic communication and inference framework between a teacher and a student agents. We first characterized semantic information on high order simplical complexes, and minimized its simplicial structure by Hodge Laplacians. We developed a masked \ac{scae} model to recursively predict query data from the semantic structures. Furthermore, a joint semantic communication and inference scheme is developed to best leverage the teacher and student's knowledge. To account for wireless communications, channel information was used in training the \ac{scae} to improve performance in poor channel. Experiments in querying citations of a coauthorship data show that the proposed methods largely reduce the simplical complex structure by up to 85\%, while improving query accuracy by 25\% and 15\% compared to baselines considering local and remote based queries. 

The proposed framework examines the potential of simplicial complexes in studying semantic information and communication, leveraging minimally sufficient semantic structure to generate query data with significantly reduced transmissions. It can be extended to learning the simplicial complexes directly from unstructured data in other modalities. Furthermore, the generation of latent simplical complexes and message passing between different orders is an important research direction. 

\bibliographystyle{IEEEtran}
\bibliography{ref}

\end{document}